# Spatially controlled electrostatic doping in graphene *p-i-n* junction for hybrid silicon photodiode

T. T. Li[1], D. Mao[1], N. W. Petrone[2], R. Grassi[3], H. Hu[4], Y. Ding[4], Z. Huang[5], G.-Q. Lo[6], J. C. Hone[2], T. Low[3], C.W. Wong[7] and T. Gu[1*]

[1]Department of Electrical and Computer Engineering, University of Delaware, Newark, DE 19711
[2]Department of Mechanical Engineering, Columbia University, New York, NY 10027
[3]Department of Electrical Engineering, University of Minnesota, Minneapolis, MN 55455
[4]DTU Fotonik, Technical University of Denmark, DK-2800 Kgs. Lyngby, Denmark
[5]Hewlett-Packard Laboratories, 1501 Page Mill Rd., Palo Alto, CA 94304
[6]The Institute of Microelectronics, 11 Science Park Road, Singapore Science Park II, Singapore 117685
[7]Mesoscopic Optics and Quantum Electronics Laboratory, University of California Los Angeles, CA 90095
* Email: tingyigu@udel.edu

**ABSTRACT**

Sufficiently large depletion region for photocarrier generation and separation is a key factor for two-dimensional material optoelectronic devices, but few device configurations has been explored for a deterministic control of a space charge region area in graphene with convincing scalability. Here we investigate a graphene-silicon *p-i-n* photodiode defined in a foundry processed planar photonic crystal waveguide structure, achieving visible - near-infrared, zero-bias and ultrafast photodetection. Graphene is electrically contacting to the wide intrinsic region of silicon and extended to the *p* an *n* doped region, functioning as the primary photocarrier conducting channel for electronic gain. Graphene significantly improves the device speed through ultrafast out-of-plane interfacial carrier transfer and the following in-plane built-in electric field assisted carrier collection. More than 50 dB converted signal-to-noise ratio at 40 GHz has been demonstrated under zero bias voltage, with quantum efficiency could be further amplified by hot carrier gain on graphene-i Si interface and avalanche process on graphene-doped Si interface. With the device architecture fully defined by nanomanufactured substrate, this study is the first demonstration of post-fabrication-free two-dimensional material active silicon photonic devices.

Hybrid integration of a lower bandgap material on a large-scale silicon (Si) photonic circuit is demanded for the active components of today's optical interconnect systems [1]. The traditional Si photonics devices have the principle drawbacks, such as the low efficiency, limited bandwidth, and lack of on-chip laser [2]. Epitaxial growth of germanium or direct bonding of III-V materials on Si has been shown to be a promising route, but the high crystal quality in an active region demands a delicate control of material growth and integration technologies [3-10]. Atomic thin graphene (G), with zero bandgap and mechanically strong in-plane structure, can be easily transferred onto Si nanophotonic platform while maintaining high crystalline quality. Among various optoelectronic applications of G [11-19], integrated G photodetectors, based on simple G-metal contacts, revealed many intriguing physics [20-26] and demonstrated highly competitive performance [27-32]. However, the built-in electrical field, which is responsible for the efficient separation of the photo-generated carriers, only exists in tens of nanometers near the G-metal contact. The absence of in-plane electric field in the bulk G region, where most of the electron–hole pairs are generated, leads to the dominant photothermal or bolometric photoelectronic response with poor on-off ratio (<1) [33-34]. Dual back gate design for achieving G *p-n* junction has been reported, which provides high flexibility of controlling the intrinsic region width and as well as tuning voltage drop between the dual gates and thus the lateral built-in electric field. The responsivity of 35 mA/W was achieved at zero-bias conditions, with a 3-dB cutoff frequency of 65 GHz. However, this structure requires sophisticated fabrication processes on the transferred G, as thus not back-end-of-line fabrication compatible [35]. G contact engineering is highly desired for achieving high-speed signal integrity and scalability of the hybrid active photonic circuits [36-39]. G-semiconductor Schottky junctions exhibit exceptional photocurrent gain up to ~$10^8$ electrons per photon [23] and low dark current leakage. The responsivity of 4 A/W at 1300 nm, 1.1 A/W at 3200 nm is demonstrated based on hot carrier tunneling in G double-layer heterostructures [40]. However, the device speed has been demonstrated is nanosecond scale, limited by the slow carrier diffusion process in non-depleted semiconductor materials [23, 40-43]. Here we report a scalable device configuration of an in-plane G *p-i-n* junction comprising the advantages of the two types contacts. The carriers confined in the limited density of states in G can be depleted through the two-dimensional junction of the hybrid device. The drift current dominant carrier transportation, both in Si and G, is a key for simultaneously addressing the device's efficiency, signal integrity and RF bandwidth [44]. The demonstrated device exhibits the responsivity of 11 mA/W at 1550 nm under zero bias with a response time of 15 ps. The converted 40 GHz photoelectric signal with 5 Hz bandwidth shows a high electrical signal-to-noise ratio of 52.9 dB.

**RESULTS**

The cross-sectional schematic of the lateral *p-i-n* junction with monolayer G coverage is depicted in Fig. 1a. The subwavelength photonic crystal (PhC) waveguide defined the intrinsic region confines photons in 250 nm thick Si membrane, which is evanescently coupled to the directly contacted G layer. The photonic crytal structure is used to enhance the photoresponsivity [45]. G fully covers the intrinsic region and directly contacts both *p*- and *n*-doped regions of Si. Aluminum electrodes, through etched oxide *vias*, forms an Ohmic contact to the heavily doped regions of Si, and isolated from G. The convergence of different work functions in the *p* and *n* regions produces an in-plane built-in electric field within both G and Si (Supplementary Materials I). In absence of external bias, the built-in electric field in G can be introduced just by asymmetric source and drain contacts. The hybrid vertical-lateral junction allows effective charge collection within the device architecture with separate regions of photon absorption and carrier conduction/amplification (Fig. 1b). The photo-excited electro-hole pairs are efficiently separated by the built-in electric field. The charge collection peaks in the center of the G layer. While the Si layer thickness (250 nm) is only 5% of its width (5 μm), most of the photo-generated carriers transfer to G layer before reaching the *p*- or *n*-doped Si region. Under the bias voltage, carrier multiplication would take place along the G material and the graphene-silicon (G-Si) interface. Schematic energy diagrams of the vertical and lateral junctions are shown in Fig. 1c. The electrostatic doping from silicon homojunction induces the Fermi-level shift in G, which is negative on *n*-Si, and positive on *p*-Si. Because of the different substrate induced the Fermi level difference in G, the built-in electric field across the lateral *p-i-n* junction is formed along the G plane. Fig. 1c also depicts the optical absorption mechanism for photon energy in infrared (IR) range, with charge carrier collection through the vertical Schottky junction. The sufficiently long space charge region in the G *p-i-n* junction is achieved by directly contact between Si *p-i-n* junction and G material, which is essential for increasing the photoresponse. The sufficient depletion region provides high flexibility for optical waveguide design with minimal overlap between waveguide mode and the lossy electrodes. The hot carriers generated in G with energy higher than Schottky barrier can be emitted into Si [46]. The top view of the CMOS processed active device layout is shown in Fig. 1d. The Raman *2D* peak intensity mapping illustrates the 60 µm long G coverage on the PhC waveguide (Fig. 1e).

**Lateral carrier transportation along G *p-i-n* junction:** For further studying the carrier transport in the hybrid space charge region, micro photocurrent mapping is conducted across the *p-i-n* junction with the 532 nm 797 nW pump as shown in Fig. 2a. The normal incident laser has a spot size of ≈ 0.6 µm, with electrical current readout by matched probes on a scanning photocurrent microscopy setup (Supplementary Material II). The scanning photocurrent mapping is used to characterize the in-plane carrier transportation. In intrinsic region, G dominates the photocarrier conduction. This is observed, first, from the symmetric photocurrent profile on the G covered section (red circles in Fig. 2a). In contrast, in monolithic Si, the lower mobility of holes compared to the electrons shifts the photocurrent peak to the *p*-doped region (blue squares in Fig. 2a). Secondly, the G-dominated photocurrent transport is observed from the extended space charge region of ≈ 7.5 µm. The majority carrier determines the photocarrier conduction in the intrinsic region of the homojunction, with lateral charge collection ($\eta_{L\_pin}$) empirically fitted by $e^{-(X-X_e)^2/L_e^2-(X-X_h)^2/L_h^2}$. $L_{e/h}$ is the mean free path for the electrons/holes in the intrinsic region. *X* is the spatial location of the laser along the *p-i-n* junction, with $X_{e/h}$ defined as borders of the intrinsic region to *n/p* doped regions. At the highly doped part of the junction, as shown in Fig. 2a, the photocurrent decays exponentially, with the decay constant determined by the minority carrier diffusion length [47-48]. Through curve fitting the model to the measured photocurrent profile in the intrinsic region, we found (1) the intrinsic region in the hybrid structure expands from 5 µm to 7.5 µm; (2) the diffusion length of the holes in the hybrid structure increases roughly from 2 µm to 3.5 µm comparing to the monolithic one (Supplementary Materials IV). A detailed collection of the carrier separation and characterization of G homojunction is provided in Supplementary Materials IV&V. The doping profile of substrate Si *p-i-n* junction is measured by electrical field microscopy (converted to an absolute value of doping concentration shown as a blue solid curve in Fig. 2b). The substrate-induced electrostatic doping in G is characterized by the Raman *G* peak wavenumber (solid red squares) (Fig. 2b) [46]. The charge-induced doping variation leads to the formation of the built-in electric field of G contacting to the substrate Si *p-i-n* junction as shown in Fig. 2c. With fixed bias on *p* and *n* contacts, we also measured the back-gate dependent photon and dark current, composed of both Si and G responses (Fig. 2d). We found our device has a back-gate threshold voltage of 1.5 V and dark current is negligible below this voltage. The photocurrent profile is illustrated in the inset of Fig. 2d (red open circles), which can be decomposed into the Si transport component (blue dashed curve) and the G component (red dashed curve).

Fig. 3a shows the measured photocurrent mapping across the G *p-i-n* junction from the wavelength of 442 nm to 832 nm. The photocurrent in the PhC waveguide region of the hybrid junction can be fitted well by our empirically model and verify the independence of excitation wavelength for $\eta_{pin}$ around the region of $X \approx 0$. The visible - NIR enhancement of the G-mediated external quantum efficiency (*EQE*), the ratio of incident photons to converted electrons is described in Fig. 3b. In this heterostructure, Si is the absorber for the visible light. Efficient carrier transfer from Si to G is enabled by their band alignment (Fig. 1c). The overall external quantum efficiency ($EQE_{G-Si}$) of the junction described below provides good corroboration with our experimental data (see also Fig. 3b):

$$EQE_{G-Si} = \left[ A_{Si}(\lambda, I) \frac{G_0 R_{G-Si}}{R_{G-Si} + R_{Rec}} + A_G(I) F_e(\lambda, T_e) \right] \eta_{L\_pin}(x) \tag{1}$$

$EQE_{G-Si}$ is the product of photocarrier generation efficiency of the vertical junction (in square bracket) and charge collection efficiency (CCE) of the lateral p-i-n junction ($\eta_{L\_pin}$). Photocarrier generation efficiency is contributed by both silicon slab (the first part) and the G layer (the second part). The silicon contribution is the product of Si PhC slab absorptance ($A_{Si}$), possible gain ($G_0$) from the carrier transfer and impact ionization from Si to G [49] and interfacial charge transfer efficiency of $R_{G-Si}/(R_{G-Si} + R_{Rec})$. $R_{G-Si}$ is built-in electric field enhanced interfacial charge transfer rate from Si to G. $R_{Rec}$ is surface dominant recombination rate in Si PhC. The charge transfer efficiency depicts the efficiency of the charge contributing to the photocurrent comparing with all of the generated charge. The photo-thermionic (PTI) current [50] from G is the product of the G photon absorbance ($A_G$) and thermionic emission efficiency of photo excited hot carriers from G to Si (modified Fowler's factor, $F_e$) [51]. We firstly characterize the $\eta_{L\_pin}$ through comparing the visible band photocurrent of monolithic and hybrid devices (Supplementary Materials IV and V). The characterized $\eta_{L\_pin}$ near the PhC waveguide region (in the middle of the intrinsic region of p-i-n junction) is then used to study the sub-bandgap PTI current in G-Si junction, with pulsed NIR light excitation launched in-plane through PhC WG.

With Si as the dominant absorber in visible band, the photocurrent mapping profile is symmetric independent of the wavelength, indicating the G dominant carrier transportation across the hybrid p-i-n junction (Fig. 3a). An ≈ 10× enhancement of the *EQE* is observed comparing the hybrid to the monolithic device (Fig. 3b), benefit from the rapid photocarrier separation in vertical G-Si junction. The *EQE* spectra in the PhC WG region shows wavelength dependent oscillations, which are originated from Fabry-Perot reflections in the air gap between suspended Si membrane and the bulk Si substrate. The spectra match finite-difference-time-domain simulations of the Si photonic crystal absorption as shown in the brown curves Fig. 3b, with (dashed curve) and without (solid curve) contributions from G absorption. The incident power is kept below 1 μW throughout the spectral response calibration.

The built-in vertical electric field and the atomically abrupt heterojunction allows for efficient carrier separation of photo-excited electron-hole pairs. The transferred hot carriers across the abrupt heterojunction might be favorable for carrier multiplication. The highest internal quantum efficiency (*IQE*) of the hybrid G silicon junction is 120% (Fig. 3c). Interfacial charge transfer efficiency describes the charge transfer efficiency on G-*i* Si interface, which saturates at high optical power (Supplementary Materials IV). Charge transfer efficiency could be calculated as $N_G/N_{Si}$, where $N_{Si}$ is the number of carriers generated per second in G Si hybrid structure, which is similar in the monolithic device; $N_G$ is the number of transferred carriers from silicon into G layer per second. The calculated highest charge transfer efficiency shown in Fig. 3c is ~ 95%. Assisted by the vertical built-in electric field, the interfacial charge transfer rate ($R_{G-Si}$) is estimated to be about 100 GHz, which is similar to the value given in literature [52]. The local recombination ($R_{Rec}$) of 20 GHz is much slower than $R_{G-Si}$ [53] (details in Supplementary Materials VI). Considering the carriers transport time (with velocity of ~$10^6$ m/s across the intrinsic region of 5 μm), the response time is estimated to be 15 ps. A hydrofluoric dip is used to remove the native oxide on Si right before the G transferring to ensure the electrical contact on the interface (Materials and Methods). The hydrogen-terminated crystalline Si surface in direct contact with G is demonstrated to be absent of Fermi-level pinning near the interface [54]. At low power input, ultrafast out-of-plane carrier separation leads to dominant G carrier transportation.

**Photo-thermionic response in vertical G-Si junction:** The sub-bandgap photocurrent generation is driven by the one photon absorption induced PTI effect in G effect, described by the second term in the square bracket in Equation (1) [50]. The ultrafast hot carrier dynamics in G has been studied with metal contacts or G double-layer heterostructure [40,55]. Here, we focus on PTI on G-*i* Si heterojunction. Photothermal excited hot carriers in G and the subsequent interfacial hot carrier emission on the Schottky junction lead to a strong wavelength dependence of the photocurrent (Fig. 4a). With a good understanding of in-plan and out-of-plane carrier transportation, the photoabsorption mechanism of the NIR light can be distinguished for separate studies. For NIR light below Si bandgap but above the barrier threshold of the Schottky junction, G is the dominant absorber through interband transitions. The out-of-plane photocarriers are collected by Si through emission of photoexcited hot carriers across the Schottky barrier [51,56]. As only photocarriers with energy above the Schottky barrier height ($\phi_b$) can be emitted to silicon, $F_e$ in Equation 1 exhibits strong wavelength dependence, and the NIR photocurrent spectrum (red squares in Fig. 4a) can be fitted by the model $F_e = C\,(\hbar\omega-\phi_b)^p$ (Solid red curve in Fig. 4a). Here $C$ is a constant. $\hbar$ is the reduced Plank constant. $\omega$ is the frequency of incident light. $p = 3$ for Si. $\phi_b$ is the Schottky barrier height of G and Si ($\phi_b = \phi_{b0} - \Delta E(T_e)$), and can be reduced through hot carrier effect in G ($\Delta E(T_e)$). Under low power continuous-wave (c.w.) excitation, the Schottky barrier $\phi_{b0}$ is measured to be 0.76 eV through curve fitting (Solid red curve in Fig. 4a). The one photon absorption in G with subsequent carrier emission on G-Si interface generates photocurrent spectrum distinguished from bare Si region (blue curve in Fig. 4a). Since atomic layer thin G has a thickness thinner than the mean free path of electron-electron scattering, the incident photons with energy above $\phi_b$ can effectively generate PTI carriers with sufficient kinetic energy to be transferred into Si, similar to the internal photo-emission process [56, 57]. As the photon energy at 1550 nm is close to the Schottky barrier, most of the photo-thermionic carriers do not have enough energy to transfer into Si with low power c.w. excitation (Inset of Fig. 4a).

Pulsed laser excitation with higher peak power increase thermal energy of the electron gas, and thus reduce the effective $\phi_b$ (Fig. 4b). The pulse duration of the excitation laser is shorter than the carrier lifetime in G, allowing hot carriers with higher electronic temperature ($T_e$) to play a role in the photocurrent, before reaching the thermal equilibrium with the carbon lattice through the electron-phonon scattering. The peak power effectively coupled onto G can be upto 10 kW cm$^{-2}$, leading to estimated transient $T_e$ of 600 K [50]. Higher $T_e$ increases the portion of the hot carriers can effectively emit into G to Si (Inset of Fig. 4b). The extra hot carrier contribution with pulsed laser excitation leads to 3.2× enhancement of photocurrent (correspondent to $\Delta E(T_e)$ = 24.5 meV), compared to c.w. light with the same wavelength and average power (Supplementary Materials VII). Photocurrent enhancement factor from the pulsed laser is weakly dependent on the reverse bias, indicating hot carrier generation and amplification are independent processes in the multi-junction device. The reverse bias enhanced built-in electric field on G-doped Si can lead to possible avalanche gain (Fig. 4b). The avalanche mechanism can be attributed to the carrier multiplication process in G-*p* Si interface. The carrier multiplication in biased G is less likely, as the lateral built-in electric field is much weaker than the vertical G-doped Si interface (further discussed in Supplementary Materials VII). The reverse bias dependent photocurrent lineshape can be fitted by the avalanche multiplication model of $M = 1/(1-(V_R/V_{BD})^k)$ [30]. The breakdown voltage ($V_{BD}$) and the power coefficient $k$ are fitted to be -0.63 V and 3.2 respectively. $M$ = 4.18 is achieved as $V_R$ set at -0.5 V bias.

To better understand the photon absorption mechanism, we compare the power dependent photocurrent between monolithic and hybrid devices (Fig. 4c). The inset of Fig. 4c distinguishes one-photon absorption in G hybrid device (solid red squares) from multi-photon absorption in Si devices (open blue circles). The dashed red line is the linear fit for a hybrid sample at low optical power. The responsivity in linear region is characterized to be 11 mA/W (*EQE* = 8.8%). At higher incident power (more than 1 MW cm$^{-2}$), the photocurrent reaches the current saturation threshold in G. An opposite trend of power dependence is observed in monolithic Si devices, where the two-photon absorption leads to enhanced responsivity at higher incident power (0.055 $P_{in}^2$ + $P_{in}$, where $P_{in}$ is the input power). The polynomial relation (solid blue curve in the insert of Fig. 4c) represents the collective photoresponse from two-photon absorption in Si (the first term) and linear absorption through mid-gap defect states in Si (the second term).

Through the 4.18 avalanche gain of our device and the slow light effect in photonic crystal structure (≈ 4 times [45]), the 60 μm long hybrid PhC waveguide is promising to achieve the photoresponsivity of 183.92 mA/W (11×4.18×4 mA/W), given the $\eta_{L\_pin}$ of 20% near the waveguide region (estimated from Fig. 3a). Through reducing the intrinsic region width to 1 um, $\eta_{L\_pin}$ can be improved to be 89% according to the model. Numerical calculation shows approximately 24% optical power is absorbed by the G layer along the 60 μm long waveguide [58]. The interfacial *IQE* at the 1550 nm is derived to be 18%.

The RF bandwidth more than 50 GHz is estimated from the sum of charge transfer time on the heterojunction and 12 ps resistance-capacitance constant [59]. The estimated RF bandwidth is verified through the measured S21 (Fig. 4d), where the 3-dB cut-off frequency is well beyond 40 GHz instrumental limit [59]. With 40GHz modulation on the 1550 nm incident light, the converted RF signal has only 5 Hz linewidth of the carrier and 52.9 dB signal to noise ratio (SNR) as shown in the inset of Fig. 4d. The devices have low noise floor given the zero-dark current at zero bias operation. Compared to other G photodetectors, the PTI effect in vertical junction, in combination with low noise carrier transportation along the lateral *p-i-n* junction, lead to the highest SNR (Table 1) with zero bias operation. Higher responsivity and operation speed can be easily achieved in other structures, as the photothermal or bolometric effect leads to light intensity dependent resistance change in G. However, the on-off ratio stays poor without effective separation of carriers [28,36]. In the visible wavelength, the G-Si *p-i-n* heterojunction demonstrated a much higher efficiency comparing with the G-WSe$_2$ heterojunction (70%) [27] and G-Si heterojunction (65%) [60]. The additional avalanche process enables additional gain in our devices. The working principle is similar to the conventional separate absorption, charge, and multiplication (SACM) APD structure [61], where the separation photocarrier generation and avalanche process can maintain the high SNR with extra gain at reverse bias.

**CONCLUSIONS**

We demonstrated waveguide integrated G-Si *p-i-n* junctions with low dark current, efficient photocarrier transport and ultrafast response. Vertical type-I band alignment between G and Si enables a broad space charge region for efficient carrier separation and drift-current dominant carrier transportation in the two-dimensional junction. External quantum efficiency of Si is improved eight times on average over the visible band, through efficient separation both across and along the hybrid vertical and lateral G-Si heterojunctions. The interfacial *IQE* of 120% has been realized at visible wavelength with silicon as the absorber. In NIR region, *EQE* of 8.8% (*IQE* of 18%) can be further improved through junction design, slow light enhancement and avalanche gain. The ultrafast response time promises the RF bandwidth of the device to be larger than 50 GHz. Under zero bias operation, the converted 40 GHz RF signal has only 5 Hz linewidth and more than 50 dB SNR.

## METHODS
### Device fabrication
The horizontal/lateral *p-i-n* diode configuration in Si membrane is fabricated in CMOS foundry by ion implantation: boron for *p*-type ($5\times10^{18}$ cm$^{-3}$, mobility 70.8 cm$^2$/V-s, resistivity 8.8 mΩ-cm) and phosphorus for *n*-type ($5\times10^{18}$ cm$^{-3}$, mobility 115 cm$^2$/V-s, resistivity 5.4 mΩ-cm). The intrinsic region is lightly *p*-doped ($10^{16}$ cm$^{-3}$, mobility 1184 cm$^2$/V-s, resistivity 0.53 Ω-cm). The surface roughness of the Si is on the scale of angstrom meter. The photonic crystal structure is defined by 248 nm deep-ultraviolet photolithography, followed by reactive ion etching to produce the hole radius of 124 ± 2 nm, with a lattice constant of 415nm. Next, an oxide cover layer is deposited for the metal insulation. Vertical *vias* in top oxide layer are patterned and etched for the contact regions, followed by standard aluminum metallization for direct contact to the heavily doped Si regions. The metallization is electrically isolated from the G layer. Chemical vapor deposition (CVD) G is grown on copper foils and transferred onto substrates using standard processing procedures. Dilute hydrofluoric acid dip is used right before the G transferring, to remove the surface oxide and achieve direct contact between G and Si photonic crystal membrane. After G transfer, the samples are exposed to air several months throughout the measurement and still function well. Any extra native oxide thickness of non-G-covered exposed region is probed by Fourier transform infrared spectroscopy and compared to the regions covered by G.

### Optical measurements
In photocurrent mapping, a supercontinuum laser source (NKT photonics) covering the whole visible bandwidth is used to characterize the photocurrent and external quantum efficiency. The sample is mounted on a confocal optical microscope with a two-axis scanning mirror, with power and wavelength control (Supplementary Materials II). Vertical (z-axis) our-of-plane coupling to the sample is used for the supercontinuum photocurrent measurements. The wavelength dependent loss of the mirrors and objective are well calibrated for deriving the effective optical power coupled onto the chip. The laser spot size is less than 1 μm. Raman spectra were collected by coupling the light scattered from the sample to an inVia Raman spectrometer through a ×100 objective (Renishaw). For measuring the hot carrier effect at telecommunication bandwidth, a pulsed laser (0.5 ps duration, 20 kW cm$^{-2}$, 10 MHz repetition rate) and c.w. laser with same average power and center wavelength (1550 nm) are coupled into the PhC region for comparison (Fig. 4b).

### Numerical simulations
A three-dimensional finite-difference-time-domain method was used to calculate the light absorption in the Si photonic crystal membrane suspended on a Si substrate. The vertical spatial resolution was set at 1 nm.

### Data availability
The data used in this study are available upon request from the corresponding author.

**ACKNOWLEDGEMENTS** The authors are grateful to A. van der Zande, C. Forsythe and F. Zhao for assistance. The authors acknowledge discussions with T. Heinz, Y. Li, P. Kim, N. Li, J. C. Campbell, A. Soman, C. Santori, R. Beausoleil, T. Otsuji, and V. Ryzhii. **Funding:** T. L., D. M. and T. G. acknowledges support from AFOSR (FA9550-18-1-0300) and NASA ECF (80NSSC17K0526). H. Hu acknowledges the support from DNRF Research Centre of Excellence, SPOC (DNRF-123), and Y. D. acknowledges Danish Council for Independent Research (DFF-1337-00152 and DFF-1335-00771). The authors acknowledge support from the National Science Foundation with grants DGE-1069240 (IGERT Optics and Quantum Electronics) and CBET-1438147.

**AUTHOR CONTRIBUTION** T. T. L. and T. G. performed the experiment, analyzed the data and wrote the manuscript. D. M. performed the small signal analysis and modeling for the RF response. T. G. designed the silicon nanophotonic layout under the guidance of C. W. W.. H. H., Y. D. and Z. H. performed the high-speed testing. R.G., T.L. performed the numerical simulations of electrostatic field distribution, and T. G. performed optical simulations. N.W.P. and J. H. prepared CVD G sample. M.B.Y., G.Q.L, and D.L.K. fabricated the silicon photonic devices.

**Competing interests:** The authors declare no competing financial or non-financial interests.

**Figures and Tables**

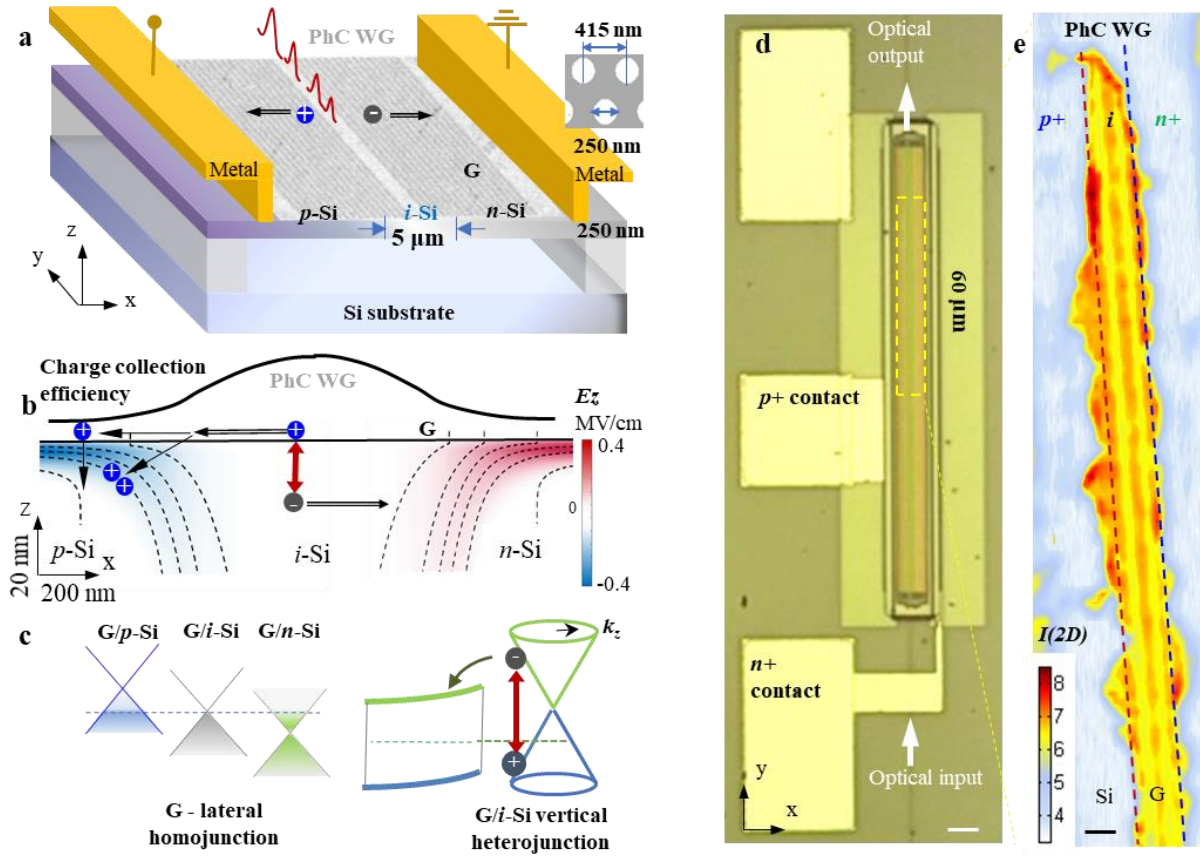

**Figure 1. Efficient carrier separation in Van der Waals contacted graphene-silicon *p-i-n* junction. a,** A schematic diagram to show the graphene-CMOS photonic crystal waveguide integration. The metal electrodes contact to the *p* and *n* sides of silicon membrane through *via*. Graphene, contacting onto the silicon photonic crystal membrane, is electrically isolated from metal electrodes. Inset: lattice constant and hole diameter of photonic crystal design. **b,** Vertical electric field near graphene-silicon interface. Arrows indicates the moving direction of carriers driven by built-in electric field. The charge collection efficiency peaks around the center of the graphene layer. **c,** The quasi fermi level of lateral homojunction and the optical absorption mechanism for photon energy in infrared range, with charge carrier collection by Schottky barriers. **d,** Top view of the device. Scale bar: 20 µm. **e.** Raman 2D peak mapping of isolated piece of single layer graphene on the intrinsic part of suspended silicon *p-i-n* junction. The two bars in the middle part reflects the substrate waveguides design. Scale bar: 3 µm.

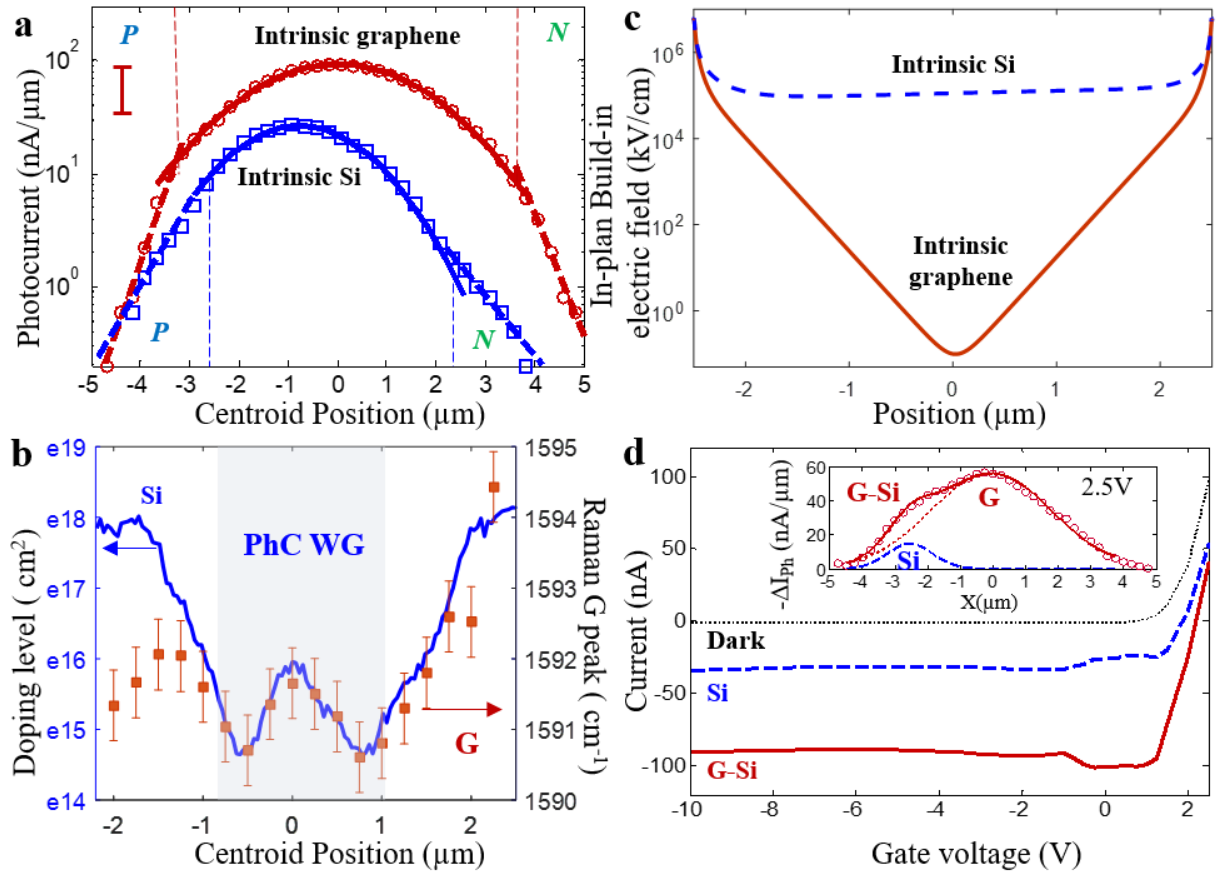

**Figure 2. Photocurrent transportation in graphene *p-i-n* junction. a,** Scanning photocurrent microscopy across the silicon *p-i-n* junction with (red circles) and without (blue open squares) graphene, under 532 nm 797 nW pump excitation without gate or external bias. Solid/dashed lines are Gaussian/exponential fits. The peak position of photocurrent profile in intrinsic region is determined by the mean free path of minority carriers. **b,** Surface conductivity scan across the G-Si junction, with the position of Raman *G* peak of graphene and substrate silicon doping level. The electrostatically formed homojunction in graphene is achieved through directly contacting to a silicon *p-i-n* junction. **c,** The calculated built-in electric field along the *p-i-n* junction ($E_x$), for monolithic silicon device (dashed blue curve) and graphene supported by silicon *p-i-n* junction. **d,** Current - gate voltage characteristics in dark (dotted black curve), as the laser spot on intrinsic silicon without (dashed blue curve) and with graphene coverage (red solid curve). Inset: the photocurrent map across the G-Si *p-i-n* junction at 2.5V gate voltage. The convoluted profile (empty circles are experiments, and the red curve is Gaussian fit) can be decomposed into graphene (dashed red curve) and silicon (blue dashed curve) contributions.

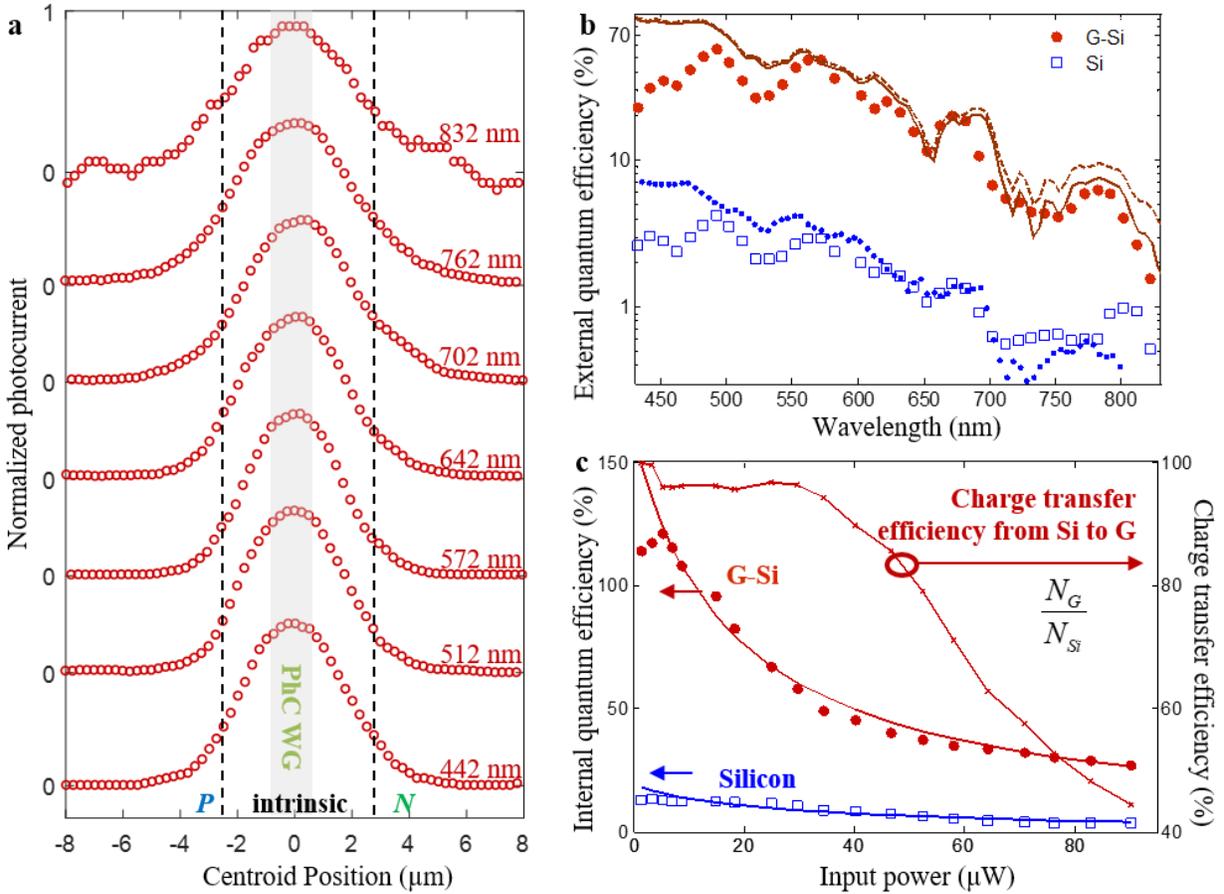

**Figure 3. Wavelength and power dependence of photocurrent map. a,** Measured Photocurrent map across the graphene *p-i-n* junction from visible to near infrared wavelength. The shadowed grey area highlights the PhC waveguide region. **b,** Measured peak external quantum efficiency (*EQE*) versus wavelength for peak photocurrent of the graphene covered (red solid circles) and silicon part (blue empty squares) of the lateral junction. The spatial photocurrent profile is same as shown in Fig. 2a. The absorption spectrum of 250nm thick silicon photonic crystal structure is plotted in red solid curve. Single layer graphene would add 2% absorption (brown dashed curve). The dotted blue line is the theoretical prediction of *EQE* of intrinsic silicon photonic crystals without absorption. **c,** Quantum efficiency for vertical junction of graphene-intrinsic silicon (brown solid circles), compared to the monolithic structure (blue empty squares). The internal emission efficiency from intrinsic silicon to graphene plotted in solid dot curve. The efficiency decreases at high optical power due to current saturation in graphene.

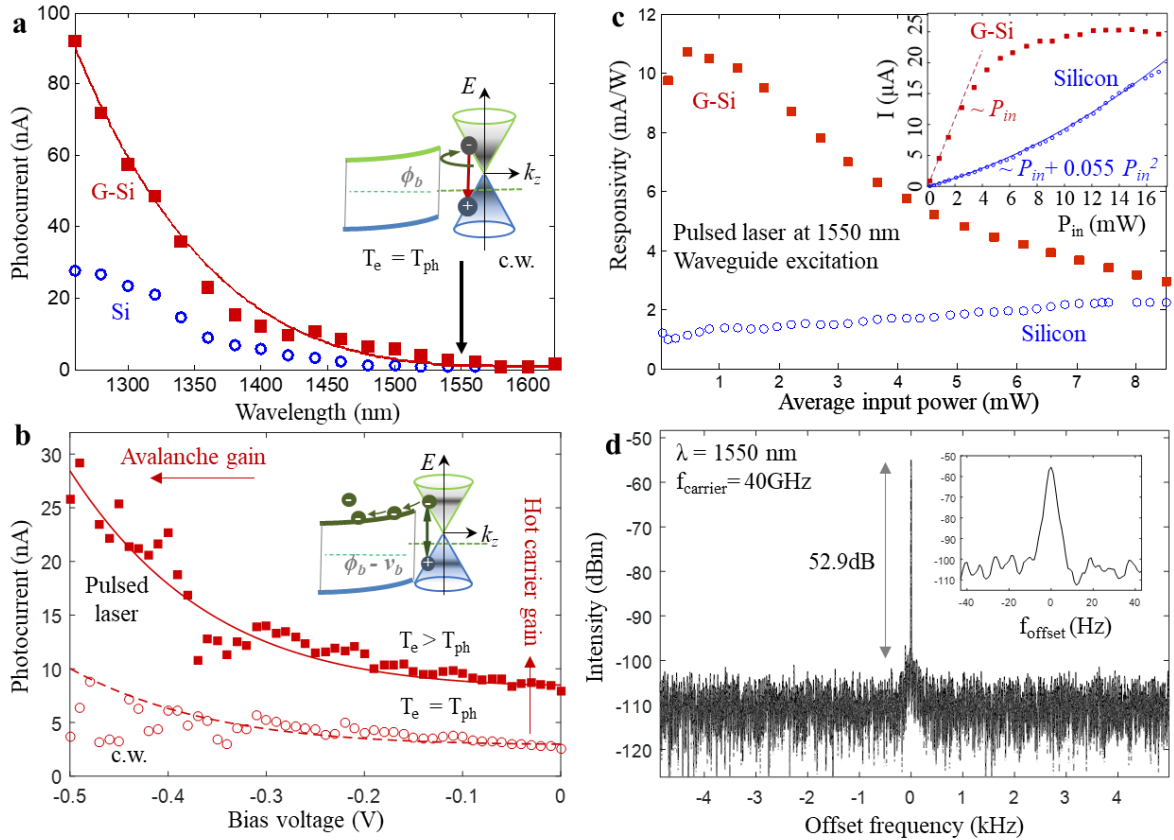

**Figure 4. Photo-thermionic (PTI) effect in vertical graphene-silicon heterojunction. a,** Low power c.w. photocurrent spectrum of PTI process. Red squares and blue circles represent hybrid and monolithic devices. **b,** Reverse bias voltage-dependent photocurrent characteristics of for graphene-Si *p-i-n* junction under c.w. laser (empty circles) and subpicosecond pulsed laser (solid squares) illumination centered at 1550nm. Inset: band diagram of PTI on graphene-Si interface and avalanche gain in Si. **c,** Responsivity versus laser pulse power for silicon (red squares) and graphene (blue empty dots) samples with zero bias. The pulsed laser propagating in intrinsic region (PhC waveguide) is centered at 1550nm. Inset: Photocurrent versus input power for devices with (red squares) and without (empty blue circles) graphene with in-plane excitation under zero bias. The dashed line is a linear relationship to the hybrid device, showing PTI dominant process at 1550 nm. The blue curve is a polynomial fit to the silicon device. **d,** Microwave spectrum of the converted photoelectric signal with a center frequency of 40 GHz. Inset: the zoom-in spectrum of the center peak, with 5Hz bandwidth and electrical signal-to-noise ratio of 52.9 dB.

**Table 1**. Performance matrix of graphene based photodetectors at room temperature

| Device Configuration | Scalability/Fabrication | IQE | SNR | Response time | SCR in G |
|---|---|---|---|---|---|
| G-metal heterojunction [36] | N/EX | 87% (NIR) | 0.64 dB | 3 ps* | -- |
| Graphene-WSe$_2$ heterojunction [27] | N/EX | >70% (759 nm) | 2.2 dB | 5.5 ps** | VJ |
| G-metal heterojunction [29] | Y/CVD | | 16 dB | 13.1 ps* | -- |
| G-metal heterojunction [28] | Y/CVD | 60% (1550 nm) | 25 dB | 24 ps* | -- |
| G p-n homojunction [35] | N/EX | High | >25 dB | 15 ps* | 80 nm |
| G/Si *p-i-n* heterojunction (this work) | Y/CVD | 18% (1550 nm) 120% (532 nm) | 52.6 dB | 15 ps* | >5 μm |
| Si *p-i-n* homojunction (control sample) | Y | NA | High | 60 ps** | -- |
| Ge/Si SACM APD [61] | Y | 16% (1300 nm) | 20 dB | 80 ps** | -- |
| G-Si heterojunction [60] | Y/CVD | 65% (Visible) | High | 1ms** | VJ |
| G: graphene; Si: silicon; SNR: signal-to-noise ratio; SCR: Space Charge Region; EX: exfoliation; CVD: Chemical vapor deposition; Graphene absorption mechanism is meant for near-infrared light; VJ: vertical junction; NIR: Near-infrared. ||||||

* RC constant limited.  ** Carrier transit time limited.

## Supplementary Materials for
**Spatially controlled electrostatic doping in graphene *p-i-n* junction for hybrid silicon photodiode**

T. T. Li[1], D. Mao[1], N. W. Petrone[2], R. Grassi[3], H. Hu[4], Y. Ding[4], Z. Huang[5], G.-Q. Lo[6], J. C. Hone[2], T. Low[3], C.W. Wong[7] and T. Gu[1*]

[1]Department of Electrical and Computer Engineering, University of Delaware, Newark, DE 19711
[2]Department of Mechanical Engineering, Columbia University, New York, NY 10027
[3]Department of Electrical Engineering, University of Minnesota, Minneapolis, MN 55455
[4]DTU Fotonik, Technical University of Denmark, DK-2800 Kgs. Lyngby, Denmark
[5]Hewlett-Packard Laboratories, 1501 Page Mill Rd., Palo Alto, CA 94304
[6]The Institute of Microelectronics, 11 Science Park Road, Singapore Science Park II, Singapore 117685
[7]Mesoscopic Optics and Quantum Electronics Laboratory, University of California Los Angeles, CA 90095
* Email: tingyigu@udel.edu

# I Theoretical potential profiles of the silicon-graphene vertical heterojunction and lateral homojunction

Let $\varphi_g$ and $\chi$ be the work function of intrinsic graphene and the electron affinity of silicon, respectively, so that the work function difference $\Delta$ between the two materials, i.e., the difference between the Fermi levels of the two isolated materials, can be expressed as

$$\Delta = \Phi_g - \chi - \Phi_n, \quad [\text{S--1}]$$

where $\varphi_n$ is the difference between the conduction band edge and Fermi level in silicon (see Fig. S1a). The latter is determined by the doping concentration. In the non-degenerate limit:

$$\phi_n = \begin{cases} kT \ln(N_c/N) & \text{for n-type doping}, \\ E_g - kT \ln(-N_v/N) & \text{for p-type doping}, \end{cases} \quad [\text{S--2}]$$

where $k$ is Boltzmann's constant, $T$ the temperature, $E_g = 1.12$ eV the energy band gap, $N = N_D - N_A$ the net doping density, and $N_c$ and $N_v$ are the effective density of states in the conduction and valence band, respectively:

$$N_c = 2 \left( \frac{2\pi m_n kT}{h^2} \right)^{3/2}, \quad [\text{S--3}]$$

$$N_v = 2 \left( \frac{2\pi m_p kT}{h^2} \right)^{3/2}. \quad [\text{S--4}]$$

Here $h$ is Planck's constant and $m_n$ and $m_p$ are the density-of-states effective masses for electron and holes, respectively. We use the values $m_n = 1.182 m_0$ and $m_p = 0.81 m_0$ ($m_0$ is the free electron mass). Also, we take $\Phi_g - \chi = 0.5$ eV. Since $\Phi_g - \chi \sim E_g/2$, it follows from [S–1]–[S–2] that, for large enough doping values $|N|$, $\Delta > 0$ ($\Delta < 0$) for n-type (p-type) doping. When the two materials are put in contact, electrons or holes are transferred from silicon to graphene depending on the sign of $\Delta$, resulting in a surface charge on the graphene layer and a band bending on the silicon side. The position of the Dirac point energy $E_d$ with respect to the Fermi level $E_F$ and the amount of band bending $\psi_s$ (see Fig. S1b) are related to each other by:

$$E_F - E_d = \Delta - \psi_s \quad [\text{S--5}]$$

With the above assumption on the sign of $\Delta$ with respect to the sign of $N$, the charge on the graphene layer is negative (positive) for n-type (p-type) silicon and a depletion region is always formed on the silicon side of the junction. Using the full-depletion approximation, i.e., assuming that the charge density $\rho = q(p - n + N)$ in silicon is given by

$$\rho(z) = \begin{cases} 0 & \text{if } z < -z_d, \\ qN & \text{if } -z_d < z < 0, \end{cases} \quad [\text{S--6}]$$

where $q$ is the electronic charge, $N = N_D - N_A$ the net doping density, $z = 0$ the position of the silicon-graphene interface, and $z_d$ the width of the depletion region, the solution of Poisson's equation $-d^2\phi/dz^2 = \rho/E_s$ is

$$\varphi(z) = \begin{cases} \varphi(-z_d) & \text{if } z < -z_d, \\ \varphi(-z_d) - \frac{qN(z+z_d)^2}{2\epsilon_s} & \text{if } -z_d < z < 0, \end{cases} \qquad [\text{S-7}]$$

with $E_s$ the dielectric constant of silicon, from which

$$\psi_s = -q[\varphi(0) - \varphi(-z_d)] = \frac{q^2 N z_d^2}{2\epsilon_s}. \qquad [\text{S-8}]$$

Since the charges on graphene and silicon must compensate each other, i.e., $Nz_d = n_g$, where $n_g$ is the net electron sheet density on the graphene layer, we get

$$\psi_s = \frac{q^2 n_g^2}{2\epsilon_s N}. \qquad [\text{S-9}]$$

In turn, $n_g$ is related to the energy difference $E_F$-$E_d$ through the graphene density of states. Using the $T = 0$ approximation:

$$n_g = \text{sgn}(E_F - E_d) \frac{(E_F - E_d)^2}{\pi(\hbar v_F)^2}. \qquad [\text{S-10}]$$

Combining [S–5] and [S–9]–[S–10], we finally get:

$$\psi_s - \frac{q^2(\Delta - \psi_s)^4}{2\epsilon_s N \pi^2 (\hbar v_F)^4} = 0, \qquad [\text{S-11}]$$

which can be solved for $\psi_s$.

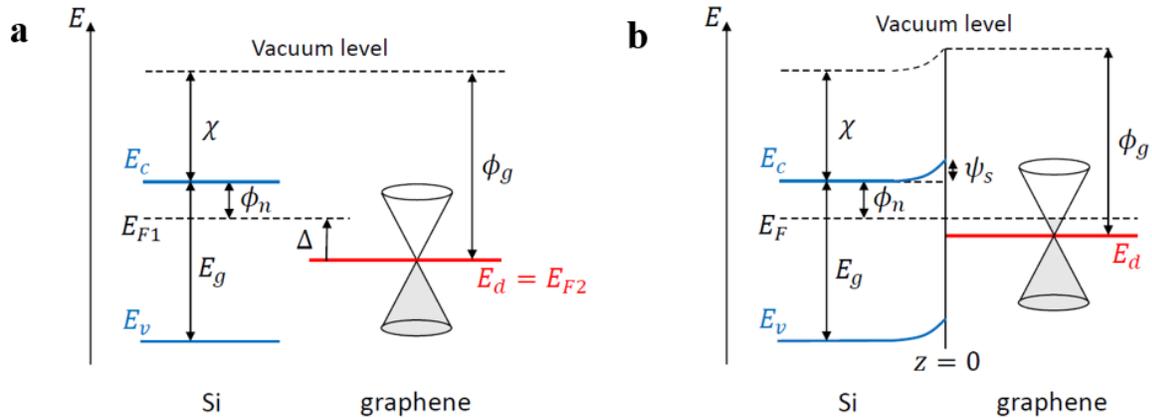

**Figure S1. Schematic band diagram of doped silicon and intrinsic graphene. a,** when the two materials are isolated and **b,** upon formation of the contact.

Fig. S2 shows the band diagram computed from the numerical solution of [S-11] for $N = -5 \times 10^{18}, 5 \times 10^{18}$ cm$^{-3}$ at T = 300 K. In the presence of a residual chemical doping $n_0$ in the graphene layer, the considerations leading from [S-1] to [S-8] are still valid. However, charge neutrality now imposes $Nz_d = n_g - n_0$, so that [S-9] is replaced by

$$\psi_s = \frac{q^2(n_g - n_0)^2}{2\epsilon_s N} \qquad [\text{S-12}]$$

And [S-11] by

$$\psi_s - \frac{q^2}{2\epsilon_s N}\left[\text{sgn}(\Delta - \psi_s)\frac{(\Delta - \psi_s)^2}{\pi(\hbar v_F)^2} - n_0\right]^2 = 0. \qquad [\text{S--13}]$$

Fig. S2 shows the band diagram computed with $n_0 = -1\times 10^{12}$ cm$^{-2}$ (p-type doping).

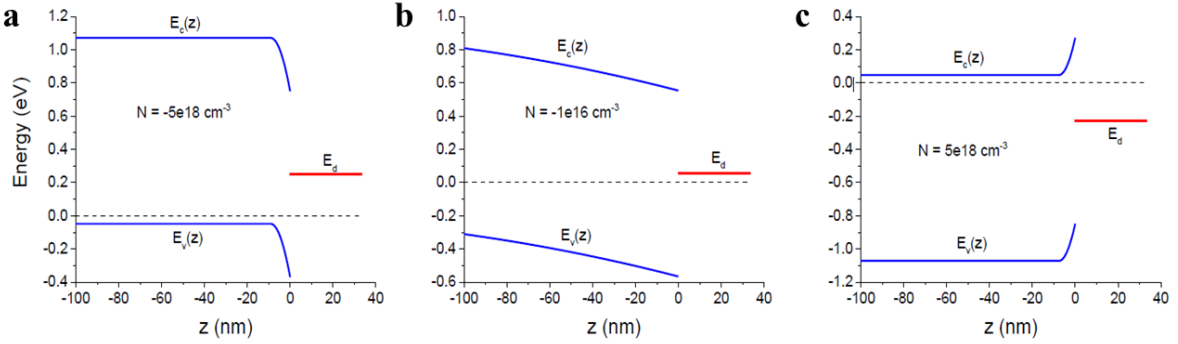

**Figure S2. Band diagram of silicon-graphene junction for a,** $N = -5\times 10^{18}$, **b,** $N = -1\times 10^{16}$, **and c,** $N = 5\times 10^{18}$ cm$^{-3}$. $E_F$ is taken as the energy reference.

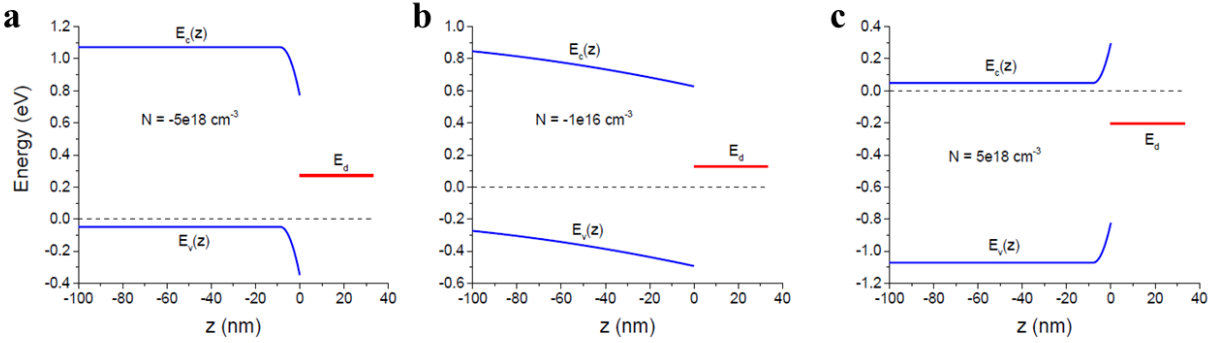

**Figure S3. Same as in Fig. S2 but for a finite p-type residual doping of graphene of $n_0 = 1\times 10^{12}$ cm$^{-2}$.** Compared to Fig. S2, the quantity $E_F$, $E_d$ is slightly decreased (which means a more p-type character of graphene, as expected), an effect which is mostly evident for low $|N|$.

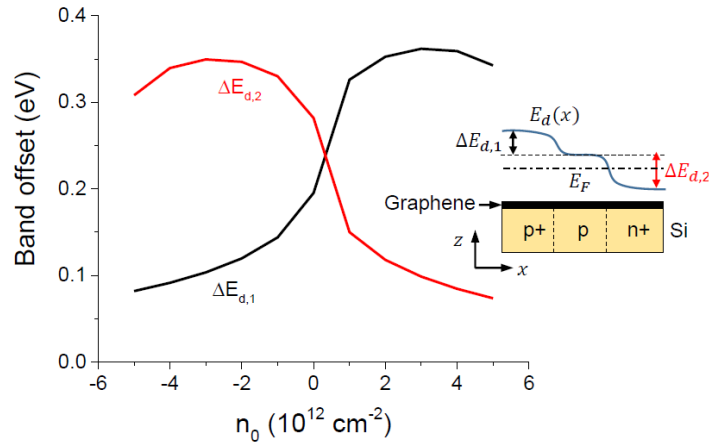

**Figure S4. Graphene band offsets in a $p+$–$p$–$n+$ silicon-graphene heterostructure.** The doping values for silicon are the same as in Fig. S2–S3. The inset shows a schematic of the device structure and the definition of $\Delta E_{d,1}$ and $\Delta E_{d,2}$ with reference to the Dirac point energy profile $E_d(X)$.

For a $p^+$-$p$-$n^+$ silicon-graphene heterostructure as the one shown in the inset of Fig. S4, the shape of the band diagram along the vertical direction $z$ at fixed longitudinal position $x$ is similar to the plots in Fig. S2–S3. The different doping values of silicon give rise to band offsets $\Delta E_{d,1}$, $\Delta E_{d,2}$ in the graphene layer (see inset of Fig. S4 for symbol definition). Fig. S4 shows the dependence of $\Delta E_{d,1}$ and $\Delta E_{d,2}$ on the residual doping $n_0$.

## II. Scanning photocurrent microscopy and integrated waveguide photocurrent measurement

Scanning photocurrent microscopy (SPCM) is used to investigate the optoelectronic properties of the hybrid G-Si structure in the visible band. SPCM provides the spatially resolved information about the photocurrent generation and transport process. In this study, the normal incident laser is focused down to a sub-micrometer full-width at half-maximum spot through a ×10 objective, with spot position controlled by a two-axis scanning mirror (Scheme 1 in Fig. S5) [S1]. The wavelength of the laser source can be tuned from visible to mid-infrared, but the operating wavelength is limited to be in the visible band by the mirrors, attenuators, and objectives.

The measured SPCM image, superimposed on the reflection image, is shown in the inset of Fig. S5. The area covered by the graphene is marked by the dashed line (illustrated in detail in Fig. S5 inset), which partially cover the silicon waveguide (solid line). Excitation light from lasers (continuous laser, tunable from 1320 to 1620nm, and subpicosecond pulsed laser, centered at 1550nm) is coupled on and off the chip through inverse tapers and lensed fiber. The photocurrent is collected by contacting standard ground-signal-ground (GSG) probe to the device with 100 μm pitch-to-pitch distance.

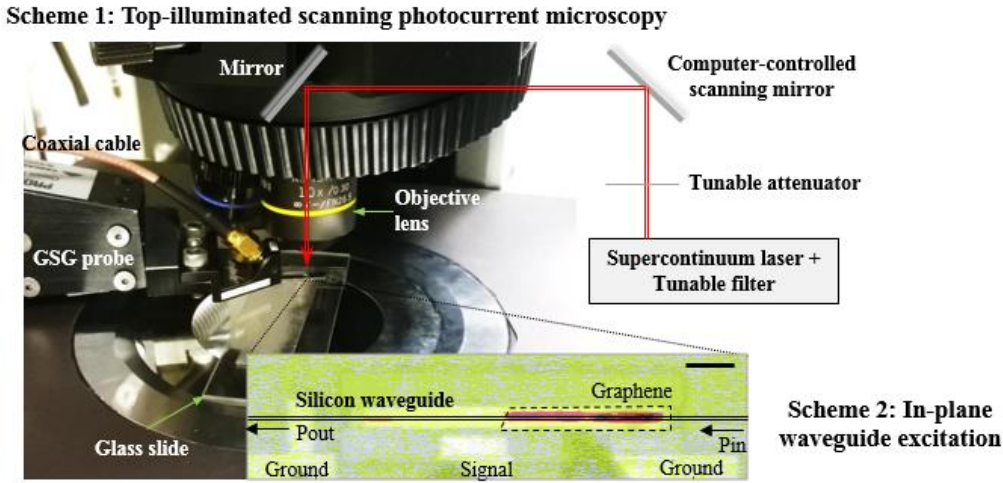

**Figure S5. Scanning photocurrent microscopy set up.** Optical image showing the top illumination and the electrical current readout, with the schematic of the optical path. Inset: the SPCM image (532nm) of the graphene covered the right part of the *p-i-n* junction. The photocurrent (red) and the reflection (grey) images are superimposed and show photocurrent generation from the graphene covered part of intrinsic region. Scale bar: 30μm. The position of silicon waveguide is indicated on the Scanning photocurrent microscopy image, with in-plane excitation ($P_{in}$) and output ($P_{out}$).

## III. Characteristics of graphene *p-i-n* junction

Lateral potential gradient along the graphene plane and low carrier densities are important for studying the built-in electric field driven photoresponse in graphene. The unsymmetrical doping in silicon leads to built-in potential in silicon substrate, which is directly contact to graphene: $V_0 = (k_BT/e)ln(N_aN_d/n_i^2 - \Delta E_F/e)$ where $k_BT/e = 0.0259$ V at room temperature, with Boltzmann constant $k_B$, room temperature $T$ and single electron charge $e$. The doping densities on both $p$ and $n$ sides are $N_a = N_d = 5 \times 10^{18}$ cm$^{-3}$. At the intrinsic region, the net carrier density in Si is $4.55 \times 10^{11}$ cm$^{-3}$, with an electrostatic potential ($V_0$) of 0.84 eV in dark. Incident pump photons are focused to a sub-1 μm$^2$ area and normally incident to the Si photonic membrane, exciting electron-hole pairs in the intrinsic region. The ultrafast carrier transfer on vertical G-Si heterojunction significantly suppress the carrier loss channels through local recombination in Si. The transferred photocarriers would be drained through the lateral built-in electric field.

In absence of external bias, the built-in electric field in graphene can be introduced by asymmetric source and drain contacts. Here the asymmetric source and drain contacts are replaced by the Si with different doping types, rather than gate activation. This configuration is illustrated with the equivalent circuit model shown in Fig. S6a. The two diode symbols correspond to $p$-$i$ and $i$-$n$ silicon junctions. Graphene (golden part) is on the top of silicon. Two tunneling junctions to the highly-doped Si on the sides (dash lines on the two sides) are formed and one Schottky contact is formed with intrinsic silicon part. In the photocarrier separation region (Graphene on intrinsic silicon of PhC WG), the Schottky barrier is calculated to be around 0.6 eV (Supplementary file S1) for near infrared photocarrier generation. The lateral built-in electric field formed in graphene (Figure S4) drives the hole towards the graphene-$p$-Si contact region, where the Schottky barrier is less than 0.05 eV. Driven by the strong vertical built-in electric field, holes travel back into the $p$-Si through field emission.

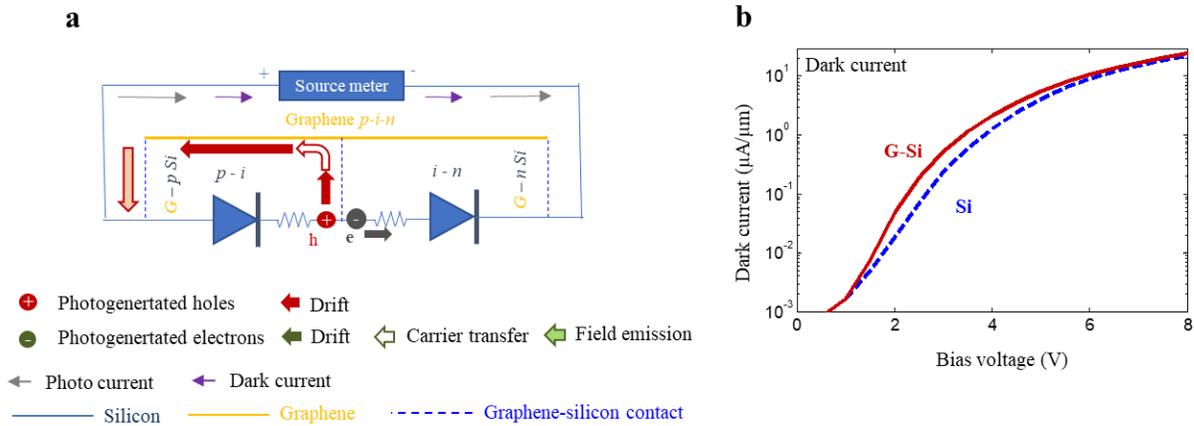

**Figure S6. Characteristics of the hybrid device. a,** Small signal model for the multi-junction formation in the hybrid $p$-$i$-$n$ junction. **b,** Measured current-voltage characteristics for graphene integrated (red solid curve) and monolithic Si $p$-$i$-$n$ junctions (blue dashed curve).

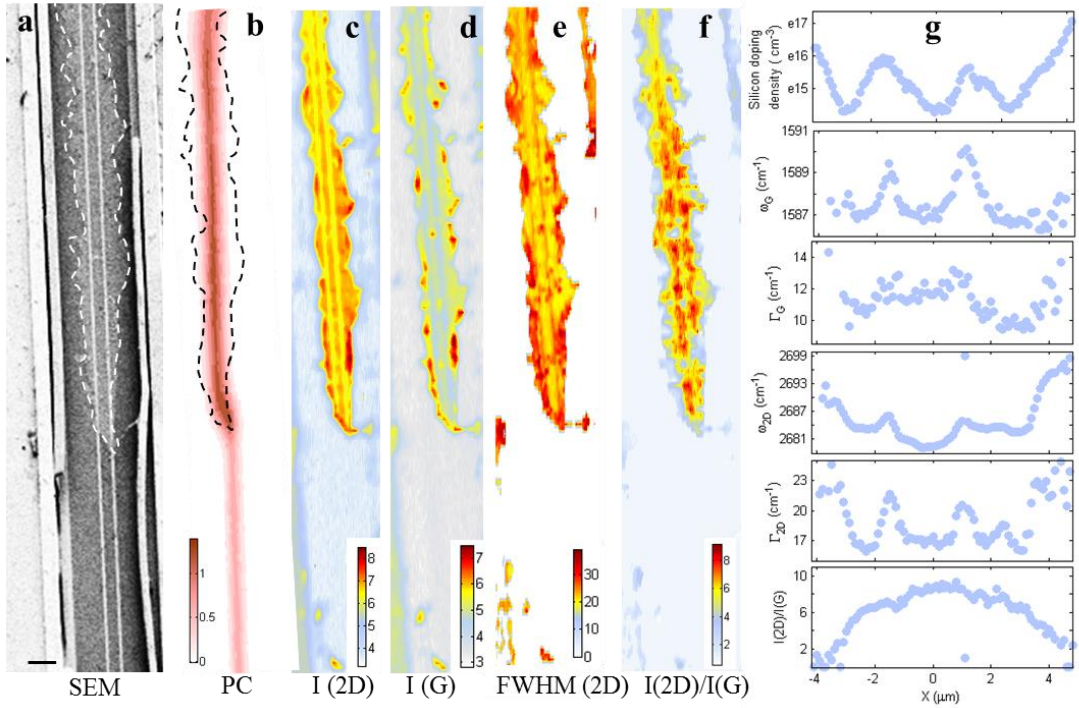

**Figure S7. Spatially-resolved Raman mapping of the graphene on silicon *p-i-n* junction. a,** SEM image of the active region covered by graphene (dashed line). **b,** Photocurrent mapping. **c,** Raman mapping of *2D* peak intensity. **d,** *G* peak intensity. **e,** Full-width-half-maximum (FWHM) of the *2D* peak. **f,** *2D* versus *G* peak ratio for the highlighted section in **a**. **g,** Doping profile of the Si substrate measured by EFM, *G* peak wavenumber ($\omega_G$), full-width half-maximum (FWHM) of *G* peak ($\Gamma_G$), *2D* peak wavenumber ($\omega_{2D}$), FWHM of *2D* peak ($\Gamma_{2D}$), and the intensity ratio of the *2D*-to-*G* peak near the intrinsic region. The patterns are repeatable at different positions along the junction.

    For the device shown in Fig. 1d, the corresponding characteristics by SEM, spatially-resolved photocurrent and Raman mapping are shown in Fig. S7a-f. The spatially dependent Raman characteristics of graphene on the Si *p-i-n* junction are compared to electrical force measurement (EFM) in Fig. S7g. The substrate screening effect from the double photonic crystal waveguide is reflected by the twin peaks in EFM mapping and the Raman *G* peak wavenumber, representing the line defect on photonic crystal plane. The substrate doping influences the wavenumber, the full-width half-maximum of the Raman *G* peak, and the intensity ratio of the *2D*-to-*G* peak.

## IV. Absorption saturation in monolithic structure

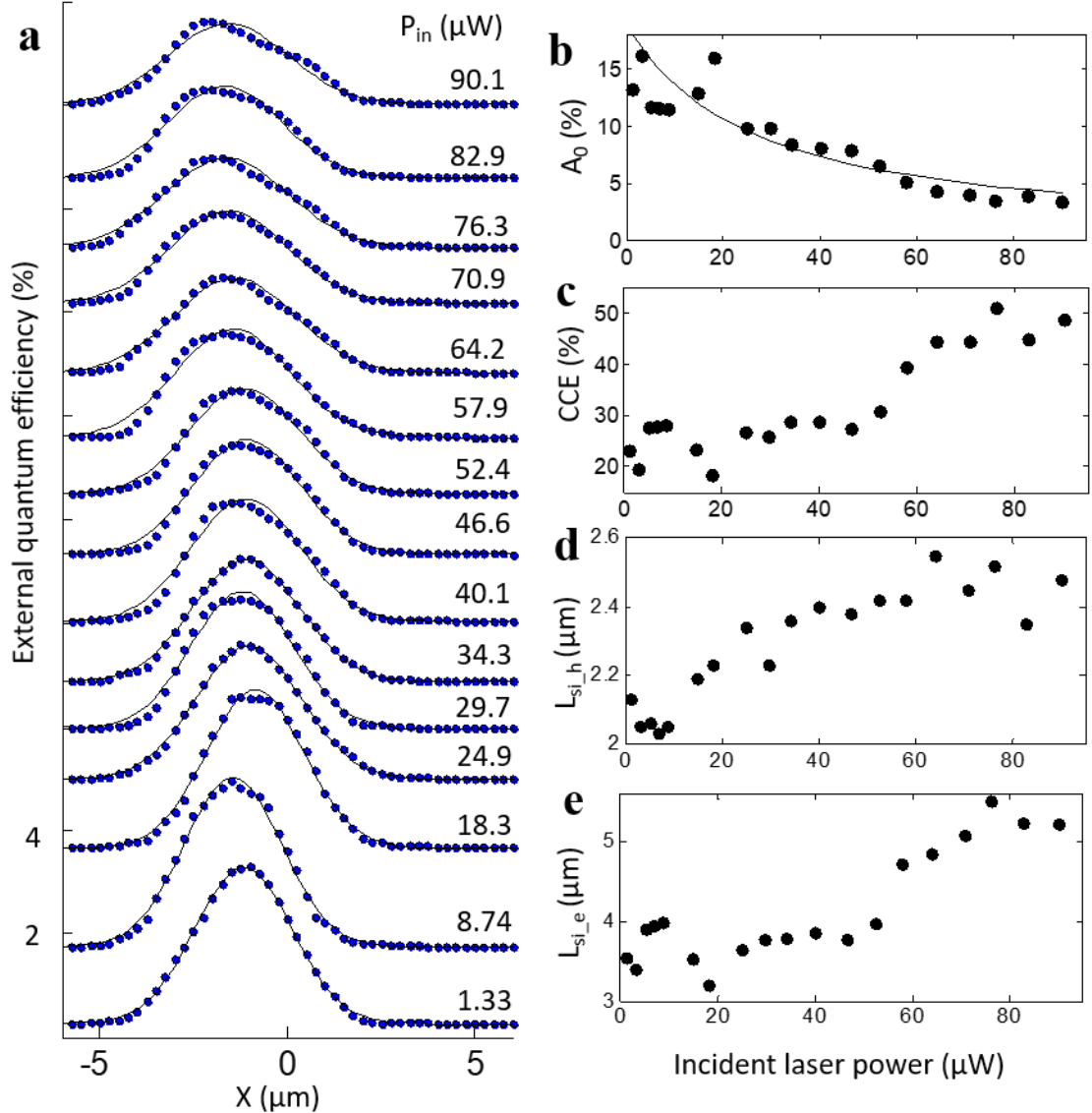

**Figure S8. Photocurrent mapping across the monolithic silicon *p-i-n* junction. a,** EQE mapping as the 532 nm green laser spot moves across the *p-i-n* junction, with plotting offset with the laser power as marked in the figure for clarity. The blue circles are experimental data and the solid black curves are corresponding fits with equations S-16. The intrinsic region is defined from -2.5 μm to 2.5 μm. The laser spot diameter is 0.6 μm **b,** Absorption coefficient versus laser power. The circles are experimental data and the curve is the fit by absorption saturation model ($A=A_0/(1+P/P_0)$). The amplitude $A_0$ is fitted to be 0.19, and saturation power $P_0$ is 25 μW. **c,** Peak charge collection efficiency of the lateral *p-i-n* junction. **d,** Mean free path length of holes and **e,** electrons in the intrinsic region of the nanostructured Si as the temperature risen with high optical injection.

The photocarrier density dynamics can be described by the master equation:

$$\frac{dn}{dt} = \frac{I\alpha}{\hbar\omega} - \frac{n}{\tau_t} \qquad [\text{S-14}]$$

where $n$ is the electron density. The first term on the right represents local carrier generation rate ($\Gamma=I\alpha/\hbar\omega$). The absorption coefficient $\alpha$ is determined by Si photonic crystals in the visible band and graphene in the infrared range. The carrier lifetime depends on carrier loss rate through local recombination ($1/\tau_{rec}$) and carrier extraction rate ($1/\tau_{transport}$): $1/\tau_t = 1/\tau_{rec} + 1/\tau_{transport}$. By solving the equation [S-14] in steady state, the density of the total carrier is $n=I\alpha\tau_t/\hbar\omega$.

At higher optical injection intensities, the local carrier density increases and reduce the absorption coefficient, both in Si and graphene:

$$\alpha(n) = \frac{\alpha_0}{1+n/n_s} = \frac{\alpha_0}{1+I/I_s} \qquad [\text{S-15}]$$

where $\alpha_0$ is the absorption coefficient at low light intensity, $n_s$ is the saturation carrier density and $I_s$ is the saturation light intensity.

We measured the photocurrent profile for monolithic devices at different optical injection levels, fitted with the equation (Fig. S8a):

$$EQE_{Si}(X) = A_{Si}(I) \times e^{-(X-X_e)^2/L_e^2 - (X-X_h)^2/L_h^2} \qquad [\text{S-16}]$$

The efficient carrier extraction ($EQE$) leads to higher saturation threshold of light intensity. $L_{e/h}$ is the mean free path for electrons/holes of majority carriers in intrinsic graphene on intrinsic Si substrate. $X$ is the spatial location of the laser, as $X_{e/h}$ are defined as the border of the intrinsic region to p/n doped regions. $I_s$ is fitted with the light intensity dependent absorption coefficient, to be 25 µW/µm² (Fig. S8b). The charge collection efficiency of lateral p-i-n junction ($\eta_{pin} = e^{-(X-X_e)^2/L_e^2 - (X-X_h)^2/L_h^2}$) is fitted as in Fig. S8c. The asymmetric mean free path of majority carriers is given in Fig. S8d-e.

## V. Current saturation in graphene

The drift current from the electron in G-Si ($I_{G-Si}$) can be expressed as [S2, S3]:

$$I_{G-Si} = \frac{W}{L}\int_0^L (n_G \mu_{G\_e} \nabla E_D + (n_{Si} - n_G)\mu_{Si\_e} \nabla E_C)dx \qquad [\text{S-17}]$$

where the drift current is the integral of electron and hole densities along the channel ($L$). The cross-section of the local channel is 5 µm wide ($W$) and 250 nm thick ($d$). The potential gradient for conduction/valance bands ($\nabla E_{c/v} = F_{Si}$) is 0.17 V/µm at low optical injection region, determined by lateral band offset (Fig. 1b). The lateral built-in electric field is shared by the atomic thin graphene layer ($\nabla E_D = F_G$), forming the potential gradient of the Dirac point ($\nabla E_D$). In Si, the mobilities for electrons and holes are $\mu_{Si\_e}$=1400 and $\mu_{Si\_h}$=450 cm²V⁻¹s⁻¹. The drift carrier velocity in Si saturates as the built-in electric filed goes beyond 0.3V. This threshold voltage becomes lower (0.22 to 0.26 V) for the hybrid device. $F_G$ and $F_{Si}$ are the built-in electric field for graphene plane and Si p-i-n junction respectively. In the hybrid structure, the electron/hole numbers per second in Si ($N_{G-Si\_Si}/P_{G-Si\_Si}$) and graphene ($N_{G-Si\_G}/P_{G-Si\_G}$) is determined by vertical potential gradient on the G-Si heterojunction (Fig.1b). The photocurrent in the hybrid device ($I_{G-Si}$) can be expressed as:

$$I_{G-Si} = N_G(\mu_{G\_e} + \mu_{G\_h}) \times F_G + (N_{Si} - N_G) \times (\mu_{Si\_e} + \mu_{Si\_h}) \times F_{Si} \qquad [\text{S-18}]$$

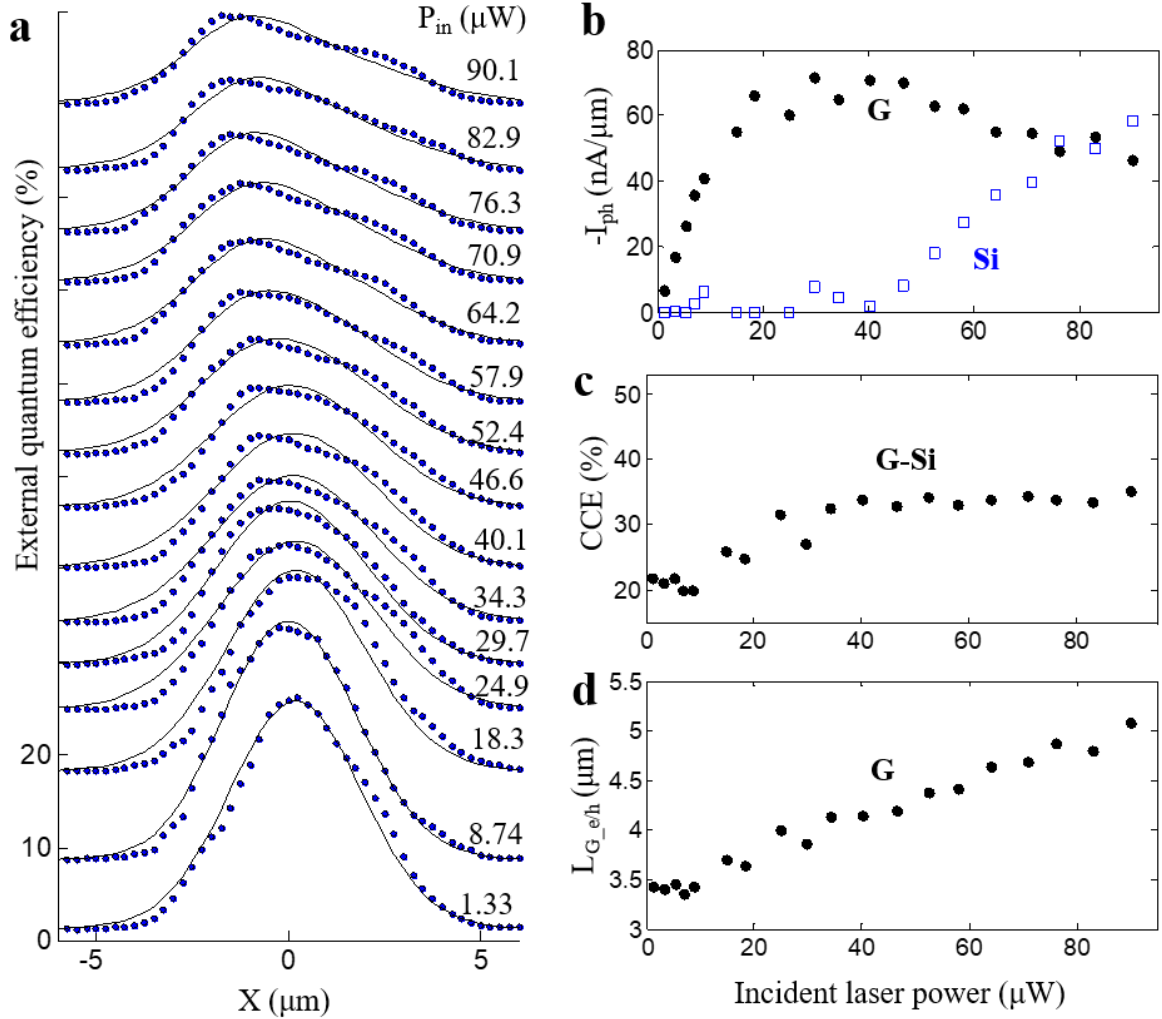

**Figure S9. Photocurrent mapping across the hybrid *p-i-n* junction. a,** External quantum efficiency (*EQE*) mapping as the 532nm green laser is spot moved across the *p-i-n* junction, with an offset of laser power (marked in the figure) for clarity. The blue circles are experimental data and the solid black curves are corresponding fits via equation S-18. The intrinsic region is defined from -2.5 μm to 2.5 μm. The laser spot diameter is 0.6 μm. **b,** Photocurrent carried by graphene (black dots) and Si (blue squares) versus laser power. **c,** Peak charge collection efficiency of the lateral *p-i-n* junction. **d,** Mean free path of holes/electrons in graphene.

The number of carriers generated per second in the graphene Si hybrid structure is same as the one in the monolithic device ($N_{Si}$). Graphene carries part of photocurrent by contacting to the Si substrate ($N_G$), determined by the vertical contact on the G-Si heterojunction (Fig. 1c). The carrier mobility in graphene is carrier density dependent [S4, S5]:

$$\mu_{G\_e/h}(n) = \frac{\mu_0}{1+(n_G/n_0)^\alpha} \qquad [\text{S-19}]$$

where $\mu_0$ = 4650 cm$^2$/Vs, $n_0$=1.1×10$^{13}$ cm$^{-2}$, and $\alpha$=2.2 at room temperature. At high optical injection region, the photocurrent profile can be fitted by:

$$I_{G-Si}(X) = I_G \times e^{-(X-X_{G\_e})^2/L_{G\_e}^2 - (X-X_{G\_h})^2/L_{G\_h}^2} + (I_{Si} - I_G) \times e^{-(X-X_{Si\_e})^2/L_{Si\_e}^2 - (X-X_{Si\_h})^2/L_{Si\_h}^2} \quad [\text{S-20}]$$

where the current in graphene ($I_G$) saturates at higher optical powers, and Si membrane carries the majority of the current ($I_{Si}$-$I_G$). By curve-fitting, the model to experimentally measured photocurrent profile across the *p-i-n* junction (Fig. S9a), the photocurrent carried by Si and graphene can be separated (Fig. S9b). Graphene carries the majority of the photocurrent until saturates at high incident laser powers (40 μW). The correspondent charge collection efficiency of the lateral *p-i-n* junction and mean free path of holes/electrons in graphene are shown in Fig. S9c,d.

## VI. Charge transfer efficiency in graphene-silicon contact

Charge separation and recombination in nanostructured *p-i-n* junction occur near-place instantaneously after the carrier generation [S4]. For photoabsorption in the visible band, the charge absorption and recommendation take place in Si. Charge transfer rate ($1/\tau_{transfer}$) and recombination rate ($1/\tau_{rec}$) determines the internal quantum efficiency of a semiconductor device: $IQE_{G-Si} = G_0(1/\tau_{tranfer})/(1/\tau_{transfer} + 1/\tau_{rec})$. Built-in electric field assisted charge transfer on the Van der Waals interface improves $IQE_{G-Si}$ in two ways: (1) introduce vertical carrier transfer channel with high charge transfer rate; (2) suppress $1/\tau_{rec}$ through reducing local carrier density. Recombination in monolithic nanostructured Si significantly reduces the quantum efficiency. The recombination current has both bulk ($R_b$) and surface contributions ($R_S$). Considering the recombination rate in bulk Si is much lower than the measured minority carrier lifetime in Si photonic crystal structures, the recombination of photogenerated carriers in the nanostructured intrinsic region is dominant by surface recombination, normalized by the surface ($S$) to volume ratio ($V$) [S5-S12]:

$$R_{Rec} = S/V \times R_S \quad [\text{S-21}]$$

The rate of surface recombination ($R_S$) is:

$$R_S = \frac{(n+n_s)(p+p_s) - n_i^2}{(p+p_s)/S_n + (n+n_s)/S_p} \quad [\text{S-22}]$$

where $n/p$ is the local electron/hole densities; $S_n$ and $S_p$ are surface recombination velocities for electrons/ holes respectively and enhanced with graphene cladding. $p_s$ and $n_s$ are the carrier density on the Si surface (in contact with graphene). In Si, the photogenerated hot carriers relax to the band edge in 300 fs. The surface recombination rate for Si nanostructure is about $2 \times 10^4$ cm/s. Given the photonic crystal structure with a lattice constant of 415 nm, hole radius of 124 nm and membrane thickness of 250 nm, the surface to volume ratio ($S/V$) is derived to be $9 \times 10^5$ /cm. As the product of surface recombination rate and surface to volume ratio, the local recombination lifetime $1/\tau_{rec}$ is estimated to be $2 \times 10^{10}$ /s (20 GHz). In the monolithic device, the local *IQE* is measured to be only 12% (blue squares in Fig. 3c), which corresponds to the $1/\tau_{transport}$ of $3 \times 10^9$ /s in lateral Si *p-i-n* junction. The hybrid structure allows much more efficient carrier transport through vertical G-Si heterojunction, with charge transfer efficiency of near 95% (red crosses in Fig. 3c). The vertical carrier transfer rate from Si to graphene is then derived to be faster than $10^{11}$/s (100 GHz).

## VII. Hot carrier separation and avalanche gain
## VII.1 Hot carrier response on G-Si junction

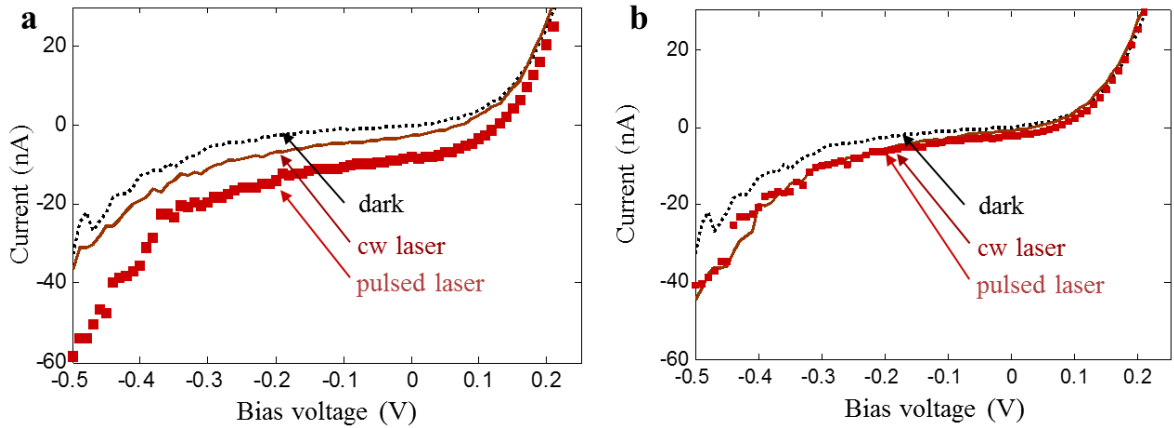

**Figure S10. The IV characteristics of the *p-i-n* junction in dark, continuous wave laser and sub-picosecond pulsed laser illumination. a,** G-Si hybrid structure. **b,** Monolithic Si structure.

In Fig. 4b, the photocurrent on the hybrid and monolithic region to near-infrared light is measured through top illumination on the same device. As shown in Fig. 1d, a 120 µm long photonic crystal waveguide is partially covered by 60 µm long graphene. Single mode fiber with a spot size of 10 µm is placed on top of the graphene covered region. The position of the fiber is fine adjusted on XYZ axis for the maximum photocurrent output. Continuous and pulsed (duration of sub-picosecond) optical signal with a center wavelength of 1550 nm is illuminated on the intrinsic region of G-Si *p-i-n* junction through the single mode fiber. The tip of the fiber is then moved to the intrinsic region for collecting the photocurrent under the same optical excitation conditions (Fig. S10). The voltage-dependent photocurrent is then derived by extracting dark current from the total current under illumination (Fig. 4b). The optical power levels under two different excitations are adjusted for generating the same photocurrent in Si *p-i-n* junction, to ensure the same average power level for pulsed and continuous wave excitation coupled to the Si photonic crystal membrane, as the hot carrier contribution to photocurrent is minimal in Si devices.

### VII.2 Discussion on carrier avalanche mechanisms

We analyzed the origin of bias dependent avalanche gain in the hybrid structure. The avalanche gain could be attributed to two junctions: (1) carrier multiplication along the biased graphene across the *p-i-n* junction. (2) avalanche on graphene-doped Si interface. Both processes might contribute to the overall reverse bias dependent photocurrent gain, as both models can be adjusted to fit the data:

*Avalanche along biased graphene:* The near-infrared light absorption in graphene generates hot carriers with the energy of half of the photon energy. The hot carriers in graphene will quickly thermalize and lose their energy through carrier-carrier scattering (in the time scale of sub 50fs [S13]) and carrier-phonon interactions (several ps) until their carrier temperature ($T_e$) reaches equilibrium as the lattice temperature ($T_{ph}$). Since only the hot carrier with energy above the Schottky barrier would be able to be collected, carrier population with higher $T_e$ leads to higher emission probability. The transient carrier temperature after sub-picosecond laser excitation is much higher than the carrier temperature in equilibrium with phonon bath (under continuous wave excitation), and thus leads to the 3.2× photocurrent enhancement through more efficient charge emission through the Schottky barrier.

As a graphene sheet is placed in an electric field ($F$) along the graphene, the charge generation,

transport, and recombination follow the 1D continuity equations and can be numerically solved [S14]. The charge generation rate through impact ionization can be approximated by the following expression:

$$U \propto T_e^{3/2} F^\alpha \exp[-(n/n_0)] \quad\quad [\text{S-23}]$$

$T_e$ is the electronic temperature, which quadratically depends on the electric field: $T_e=T_L[1+(F/F_{CT})^2]$. $T_L$ is the lattice temperature. $F_{CT}$ is a critical field for the onset of hot carrier effects, limited by various scattering mechanisms. $\alpha$ is a numerical factor of 1.7. $n$ is local carrier density. $n_0$ is around $1\times10^{12}$ cm$^{-2}$, and slightly increases with $T_e$. The multiplication factor can be derived as $M=(n+U)/n$. The clean graphene Si interface with asymmetric semiconductor contact lowers the FCT to be only 1.1kV/cm, as fitted to the experimental data in Fig. 4b.

*Avalanche on graphene-Si interface:* The voltage dependence of photocurrent has contributions from both thermionic emission (TE) and avalanche gain on G-Si interface. As the intrinsic region of Si is moderately doped ($10^{16}$ cm$^{-3}$), thermionic emission dominants the carrier transport process. The voltage dependence of TE can be expressed as $\sqrt{V_R + \phi_B/E_0} \exp(qV_R/\varepsilon')$, where $E_0$ and $\varepsilon'$ are two constants determined by the Si doping and operation temperature, and $V_R$ is the reverse bias. The avalanche multiplication factor follows the empirical model $M=1/(1-(V_R/V_{BD})^k)$ [29]. The bias dependence of photocurrent is proportional to the product of TE and $M$. The breakdown voltage ($V_{BD}$) and the power coefficient $k$ are fitted to be -0.63 V and 3.2 respectively. $M = 4.18$ is achieved as $V_R$ set at -0.5V bias. Photocurrent ratio between the pulsed laser and continuous wave excitation is weakly dependent on the reverse bias, indicating the independence of the hot carrier generation and amplification processes.

**VII.3 Dark current analysis by Landauer transport model**
As shown in Fig. S10, both the photocurrent and dark current depend on the reverse bias. The limited density of states in graphene leads unique carrier transport behavior on G-Si junction. Landauer transport formalism can be used for predicting the thermionic emission current of carrier injection on direct contact between G-Si interface [S15-S16]:

$$J_0 = \frac{2}{\pi} R_{Injection} q_0 \left(\frac{k_B T}{\hbar v_F}\right)^2 \left(\frac{\phi_0}{k_B T}+1\right) \exp\left(\frac{-\phi_0}{k_B T}\right) \quad\quad [\text{S-24}]$$

where $R_{Injection}$ is the charge injection rate, $q_0$ is the elementary charge, $k_B$ is the Boltzmann's constant, $\hbar$ is the reduced Planck's constant, $v_F$ is the graphene Fermi velocity, $\phi_B$ is the Schottky barrier at zero bias, and $V_R$ is the reverse bias. Laudauer transport model can be applied to dark current analysis on Graphene-Si contact.